\documentclass[conference]{IEEEtran}
\usepackage{nonfloat}
\usepackage{graphicx}
\usepackage{amssymb}
\usepackage{epstopdf}
\usepackage{amsmath}
\usepackage{amsthm}
\newtheorem{mydef}{Definition}
\newtheorem{lem}{Lemma}
\newtheorem{thm}{Theorem}
\newtheorem{cor}{Corollary}

\begin{document}
\title{An Upper Bound on the Convergence Time for Distributed Binary Consensus}

%
       
       \author{
\IEEEauthorblockN{Shang Shang\IEEEauthorrefmark{1},  Paul W. Cuff\IEEEauthorrefmark{1}, Sanjeev R. Kulkarni\IEEEauthorrefmark{1}, and Pan
Hui\IEEEauthorrefmark{2},}
\IEEEauthorblockA{\IEEEauthorrefmark{1}Department of Electrical Engineering,
Princeton University, Princeton NJ, 08540, U.S.A.
} \IEEEauthorblockA{\IEEEauthorrefmark{2}Deutsche Telekom
Laboratories, Ernst-Reuter-Platz 7, 10587 Berlin, Germany\\
\IEEEauthorrefmark{1}\{sshang, cuff, kulkarni\}@princeton.edu, \IEEEauthorrefmark{2}pan.hui@telekom.de}
}
       
\maketitle
\begin{abstract}
The problem addressed in this paper is the analysis of a distributed consensus algorithm for arbitrary networks, proposed by B\'en\'ezit et al.. In the initial setting, each node in the network has one of two possible states (\emph{``yes"} or \emph{``no"}). Nodes can update their states by communicating with their neighbors via a 2-bit message in an asynchronous clock setting. Eventually, all nodes reach consensus on the majority states. We use the theory of electric networks, random walks, and couplings of Markov chains to derive an $O(N^4\log N)$ upper bound for the expected convergence time on an arbitrary graph of size $N$.
\end{abstract}
\section{Introduction}
\label{sec:intro}

The problem of distributed and quantized average  consensus on networks has received considerable attention recently \cite{Boyd}\cite{Zhu}\cite{Kashyap}. It models averaging in a network with finite capacity channels \cite{Kashyap}. It is of interest in the context of coordination of autonomous agents, estimation, and distributed data fusion on sensor networks, peer-to-peer systems, etc.\cite{Boyd}\cite{Drief}. A special case of distributed averaging is the distributed binary voting problem, where all the nodes in the network vote ``\emph{yes}" or ``\emph{no}" and a majority opinion is desired \cite{Benezit}\cite{Mossel}. 

\begin{figure}[!t]
\centering
\centerline{\includegraphics[width=6.5cm]{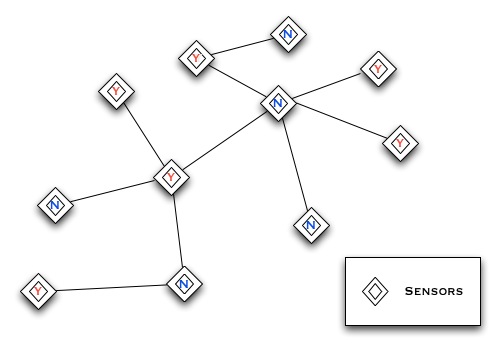}}
\caption{{\em Sensor nodes deployed in decision-making.}}
\label{sensor}
\end{figure}

Distributed algorithms requiring limited communication and computation are specially appealing to remote and extreme environments. An example to motivate this problem is sensor decision-making in a network. For example, as shown in Fig. \ref{sensor}, sensors are deployed to measure if an earthquake is happening or not in a certain region. The limited communication only happens between sensors. Each sensor has limited battery. An algorithm with good convergence speed is desired. 

This work is motivated by the distributed binary voting algorithm proposed by B\'en\'ezit et al. \cite{Benezit}. They showed that the algorithm reaches consensus on a quantized interval that contains the average almost surely. However, they did not analyze the convergence time. In \cite{Mossel}, the authors studied the convergence speed in the special case of regular graphs for a similar distributed binary consensus algorithm. Draief et al. derived an expected convergence time bound depending on the second largest eigenvalue of a doubly stochastic matrix characterizing the algorithm and voting margin \cite{Drief}, instantiating the bound with some particular networks, yet no specific bound is provided for an arbitrary graph. In this paper, we derive an $O(N^4\log N)$ upper bound for the convergence speed of the distributed algorithm for arbitrary connected graphs using results on electric networks, random walks, and Markov chain coupling .

The remainder of this paper is organized as follows. Section 2 describes the algorithm proposed in \cite{Benezit} and formulates the convergence speed problem. In Section 3, we derive our polynomial bound for this algorithm. In Section 4, we give a simple example on how to derive an upper bound given the topology of the network, simulation results are provided to justify the analysis. We provide our conclusions in Section 5.

\section{Problem Statement}
\label{sec:problem}

A network is represented by a connected graph $\mathcal{G = (V,E)}$, where $\mathcal{V} =\{1,2,...,N\}$ is the set of nodes and $\mathcal{E}$ is the set of edges. $(i,j)\in \mathcal{E}$ if nodes $i,j$ can communicate with each other. $\mathcal{N}_i$ is the set of neighbors of node $i$. 

Consider a network of $N$ nodes, labeled 1 through $N$. As proposed in \cite{Boyd}\cite{Kashyap}\cite{Benezit}, each node has a clock which ticks according to a rate 1 exponential distribution. By the superposition property for  the exponential distribution, this set up is equivalent to a single global clock with a rate $N$ exponential distribution ticking at times $\{Z_k\}_{k\ge0}$. The communication and update of  states only occur at  $\{Z_k\}_{k\ge0}$. When the clock of node $i$ ticks, $i$ randomly chooses a neighbor $j$ from the set $\mathcal{N}_i$.  We say edge $(i,j)$ is activated. Let $S^{(i)}(t)$ denote the state of node $i$ at time $t$. $S^{(i)}(t) \in \{S^+, S^-, W^+, W^-\}$, representing \emph{strong positive}, \emph{strong negative}, \emph{weak positive}, and \emph{weak negative} respectively, where $S^\pm = \pm2$ and  $W^\pm = \pm1$.  The two nodes $i,j$ then update their states according to the following update rules:

\begin{enumerate}
  \item If $S^{(i)}(t)=S^{(j)}(t), \\S^{(i)}(t+1)=S^{(j)}(t+1)=S^{(i)}(t);$
  \item If $|S^{(i)}(t)|>|S^{(j)}(t)|$ and  $S^{(i)}(t)\cdot S^{(j)}(t)<0, \\S^{(i)}(t+1)=-S^{(j)}(t), \;S^{(j)}(t+1)=S^{(i)}(t)$, and vice versa;
  \item If $|S^{(i)}(t)|>|S^{(j)}(t)|$ and  $S^{(i)}(t)\cdot S^{(j)}(t)>0, \\S^{(i)}(t+1)=S^{(j)}(t), \;S^{(j)}(t+1)=S^{(i)}(t)$, and vice versa;
  \item If $S^{(i)}(t)=-S^{(j)}(t), \\S^{(i)}(t+1)=\textrm{sign} \left(S^{(j)}(t)\right),\\S^{(j)}(t+1)=\textrm{sign}\left(S^{(i)}(t)\right).$
 \end{enumerate}

\begin{mydef}[Convergence]
A binary voting reaches convergence if all states of nodes on the graph are positive or all states are negative.
\end{mydef}

\begin{figure}[t]
\centering
\centerline{\includegraphics[width=6.5cm]{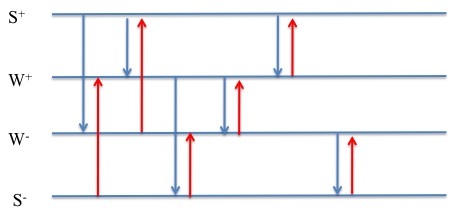}}
\caption{{\em Update rules for distributed binary vote}\cite{Benezit}. \em{The figure shows update principles: when opposite ``strong opinion"s meet, they both turn into ``weak opinion"s; ``strong opinion" affects ``weak opinion"; and swap principle.}}
\label{voting}
\end{figure}

We show the update rules in Fig. \ref{voting}. Note that this algorithm supposes that there is an odd number of nodes in the network, in order to guarantee convergence regardless of initial votings of nodes. 

Let $|S^+|$ denote the number of the \emph{Strong Positive} opinions and $|S^+(t)|$ denote the number of the \emph{Strong Positive} opinions at time $t$. A quick validation of the convergence of the algorithm in Section \ref{sec:problem}: we notice that the $S^+$ and $S^-$ will only annihilate each other when they meet, otherwise they just take random walks on the graph.  So only the majority\emph{strong opinions} will be left on the graph in the end. We also notice that \emph{strong opinions} can influence \emph{weak opinions} as shown in Fig. \ref{voting}. Eventually all agents will take the sign of the majority  \emph{strong opinions}. Because the graph has finite vertex, and this Markov chain  has finite states, thus convergence will happen in finite time almost surely. In this paper, we are interested in studying the convergence time of this distributed algorithm on an arbitrary graph.

\section{An Upper Bound on the Convergence Time for Distributed Binary Consensus}\label{main}
The main result of this work is the following theorem:
\begin{thm}
\label{mainth}
The upper bound for the binary voting algorithm is $\mathcal{O}(N^4\log(N))$.
\end{thm}
We use the analogy of electric networks and random walks to derive the upper bound. Before we proceed to the  formal proof, it is helpful to give some definitions of random walk and provide the notation we will use in the proof.

\subsection{Definition and Notation}\label{def}
\begin{mydef}[Hitting Time]
For a graph $\mathcal{G}$, let $\mathcal{H}{(i,j)}$ denote the expected number of steps a random walk beginning at $i$ must take before reaching $j$. Define the ``hitting time'' of $\mathcal{G}$ by $\mathcal{H(G)} = \max_{i,j}\mathcal{H}(i,j)$.
\end{mydef}

\begin{mydef}[Meeting Time]

Consider two random walkers are placed on $\mathcal{G}$. At each tick of the clock, they move according to some joint probability distribution. Let $\mathcal{M}{(i,j)}$ denote the expected time that the two walkers meet at the same node or they cross each other through the same edge (if they move at the same time). Define ``meeting time" of $\mathcal{G}$ by $\mathcal{M(G)} = \max_{i,j}\mathcal{M}(i,j)$.
\end{mydef}

Define a \emph{simple random walk} on $\mathcal{G}$, $\mathcal{X}_S$, with transition matrix $P^{S} = (P_{ij})$:
\begin{itemize}
\item $P^S_{ii}: =0$ for $\forall i \in \mathcal{V}$,\\
\item $P^S_{ij}: =\frac{1}{|\mathcal{N}_i|}$ for $(i, j) \in \mathcal{E}$.  
\end{itemize}
$\mathcal{N}_i$ is the set of neighbors of node $i$ and $|\mathcal{N}_i|$ is the degree of node $i$.

Define a \emph{natural random walk $\mathcal{X}_N$} with transition matrix $P^{N} = (P_{ij})$:
\begin{itemize}
\item $P^N_{ii}=1-\frac{1}{N}$ for $\forall i \in \mathcal{V}$,
\item $P^N_{ij}=\frac{1}{N|\mathcal{N}_i|}$ for $(i, j) \in \mathcal{E}$.
\end{itemize}

Define a \emph{biased random walk $\mathcal{X}_B$} with transition matrix $P^B = (P_{ij})$:
\begin{itemize}
\item $P^B_{ii}: =1-\frac{1}{N} - \sum_{k\in \mathcal{N}_i}\frac{1}{N|\mathcal{N}_k|}$ for $\forall i \in \mathcal{V}$,
\item $P^B_{ij}: =\frac{1}{N}\left(\frac{1}{|\mathcal{N}_i|} + \frac{1}{|\mathcal{N}_j|}\right)$ for $(i, j) \in \mathcal{E}$.
\end{itemize}

\subsection{The Analogy}\label{ana}
Before two opposite strong opinions $m,n$ meet each other, they take random walks on the graph $\mathcal{G}$. Their marginal transition matrices are both $P^B$. It may be tempting to think that they are taking \emph{natural random walk} as stated in \cite{Zhu} for a similar problem, but after a second close look, we find that there are two sources stimulating the random walk from $i$ to $j$, $\forall(i,j)\in \mathcal{E}$: one is the clock of the node $i$, $P^1_{ij} = P^N_{ij}$; the other one is the clock of its neighbor $j$, $P^2_{ij} = P^N_{ji}$. Thus $P_{ij} = P^1_{ij} + P^2_{ij}$, i.e., the transitional matrix is actually $P^B$ instead of $P^N$. Denote this random process as $\mathcal{X}$. Since the two random walks $m, n$ cannot move at the same time, if they are not adjacent. Suppose $m$ is at node $x$, and $n$ is at node $y$. 

For $x \notin \mathcal{N}_y$,  and  $i \in \mathcal{N}_x$, we have
\begin{eqnarray}\nonumber 
&& P_{\mathcal{X}\textrm{joint}}(m\textrm{   moves from } x \textrm{ to } i,\textrm{ } n\textrm{ does not move}) \\ \nonumber &= &
P^B_{xi}(m) - P_{\mathcal{X}\textrm{joint}}(\textrm{m moves from } x \textrm{ to } i\textrm{, } n\textrm{ moves}) \\ 
& = &P^B_{xi}.
\label{bayes}
\end{eqnarray}
Similar for $P_{\mathcal{X}\textrm{joint}}(n\textrm{   moves from } y \textrm{ to } j ,\textrm{ } m\textrm{ does not move})$.
 Also, 
\begin{eqnarray} \nonumber
&& P_{\mathcal{X}\textrm{joint}}(m\textrm{ does not move, } n\textrm{ does not move}) \\
&=& 1 - \sum_{i\in \mathcal{N}_x}P^B_{xi} - \sum_{j\in \mathcal{N}_y}P^B_{yj}.
\end{eqnarray}
For $x \in \mathcal{N}_y$ and $i \neq y$ we have,  
\begin{eqnarray}
\nonumber 
&& P_{\mathcal{X}\textrm{joint}}(m\textrm{   moves from } x \textrm{ to } i,\textrm{ } n\textrm{ does not move}) \\ \nonumber &= &
P^B_{xi}(m) - P_{\mathcal{X}\textrm{joint}}(\textrm{m moves from } x \textrm{ to } i\textrm{, } n\textrm{ moves}) \\ 
& = &P^B_{xi}.
\end{eqnarray}
\begin{eqnarray} 
 P_{\mathcal{X}\textrm{joint}}(m\textrm{ moves to y, }n \textrm{ moves to x}) = P^B_{xy}
 \label{meet}
\end{eqnarray}
\begin{eqnarray} 
\nonumber
&& P_{\mathcal{X}\textrm{joint}}(m\textrm{ does not move, } n\textrm{ does not move}) \\
&=& 1 - \sum_{i\in \mathcal{N}_x}P^B_{xi} - \sum_{j\in \mathcal{N}_y}P^B_{yj} + P^B_{xy}.
\label{stay}
\end{eqnarray}

\subsection{Meeting Time on Weighted Graph}\label{mt}
\begin{lem}
The biased random walk $\mathcal{X}_B$ is a reversible Markov process.
\end{lem}
\begin{proof}
A Markov process is said to reversible if the transition probability, $P$, between each pair of states $i$ and $j$ in the state space obey $\pi_iP_{ij} = \pi_jP_{ji}$.

Let $\pi$ be the stationary distribution of $\mathcal{X}_B$. It is easy to get
\begin{equation}
\pi_i = \frac{1}{N}
\end{equation}
 for all $i\in \mathcal{V}$.
Thus $$\pi_iP^B_{ij} = \pi_jP^B_{ji}.$$
\end{proof}

\begin{lem}\label{circle}
$\mathcal{H}_{P^B}(x, y) + \mathcal{H}_{P^B}(y, z) + \mathcal{H}_{P^B}(z, x) = \mathcal{H}_{P^B}(x, z) + \mathcal{H}_{P^B}(z, y) + \mathcal{H}_{P^B}(y, x). $
\end{lem}
\begin{proof}
A direct result from Lemma 2 in Chap 3 of Aldous-Fill's book \cite{Aldous} since $\mathcal{X}_B$ is reversible. 
\end{proof}

\begin{thm} \label{hitting}
$\mathcal{H}_{P^B}(\mathcal{G}) < \frac{N^4}{2}$.
\end{thm}
\begin{proof}
The biased random walk $\mathcal{X}_B$ defined above is a random walk on a weighted graph with edge weight
\begin{equation}\label{weight}
w_{ij}: = \frac{1}{N}\left(\frac{1}{|\mathcal{N}_i|} + \frac{1}{|\mathcal{N}_j|}\right) \textrm{ for } (i, j) \in \mathcal{E}.
\end{equation}
\begin{equation}
w_{ii} : = 1 - \sum_{j \in \mathcal{N}_i}w_{ij}.
\end{equation} 
\begin{equation}
w_i = \sum_{j\in \mathcal{V}}w_{ij} = 1, \; w = \sum_i{w_i} = N. 
\end{equation}
It is well-known that there is an analogy between a weighted graph and an electric network, where a wire linking $i$ and $j$ has conductance $w_{ij}$, i.e., resistance $1/w_{ij}$ \cite{Aldous}\cite{Nash}. And they have the following relationship
\begin{equation}\label{commute}
\mathcal{H}_{P^B}(x, y) + \mathcal{H}_{P^B}(y, x) = wr'_{xy},
\end{equation}
where $r'_{xy}$ is the effective resistance in the electric network between node $x$ and node $y$. Since the degree of any node is at most $N-1$, for $(i, j) \in \mathcal{E}$, 
\begin{eqnarray}\nonumber 
w_{ij} &=& \frac{1}{N}\left(\frac{1}{|\mathcal{N}_i|} + \frac{1}{|\mathcal{N}_j|}\right) \\ 
\nonumber & \ge& \frac{1}{N}(\frac{1}{N-1} + \frac{1}{N-1}) \\
& >& \frac{1}{N}(\frac{1}{N} + \frac{1}{N}) = \frac{2}{N^2}.
\end{eqnarray}
Thus $r_{ij} < \frac{N^2}{2}$. And consequently, $r'_{ij} \le r_{ij} <\frac{N^2}{2}$, indicating that the effective resistance between any two neighboring nodes is less than $N^2/2$.

For $\forall x, y \in \mathcal{V}$, $r'_{xy} \le (N-1)\cdot r_{ij} <  \frac{N^3}{2}$. The worst case is $N-1$ resistors connected in serial. Connecting any more resistors between any two nodes will only decrease the total resistance. By Equation (\ref{commute}), we have 
\begin{eqnarray}\nonumber 
\mathcal{H}_{P^B}(x,y) &<& \mathcal{H}_{P^B}(x, y) + \mathcal{H}_{P^B}(y, x) \\ \nonumber
&=& wr'_{xy}  \\ \nonumber
&<& N\cdot\frac{N^3}{2}  \\
&=& \frac{N^4}{2}.
\end{eqnarray}
This completes the proof.
\end{proof}
Note that this is an upper bound for arbitrary connected graphs. A tighter bound can be derived for given topology of networks. A simple example of star networks will be given in the next section. 

\begin{mydef}[Hidden Vertex]
A vertex $t$ in a graph is said to be hidden if for every other point in the graph,  $\mathcal{H}(t, v) \le\mathcal{H}(v, t)$. A hidden vertex is shown to exist for all reversible Markov chains in \cite{Coppersmith}.
\end{mydef} 

\begin{thm}\label{meeting}
The meeting time of any two opinions on the network $\mathcal{G}$ is less than $4\mathcal{H}_{P^B}{(\mathcal{G})}$.
\end{thm}
\begin{proof}
In order to prove the theorem, we construct a coupling Markov chain, $\mathcal{X}'$ to assist the analysis. $\mathcal{X}'$ has the same joint distribution as $\mathcal{X}$ in Section \ref{ana} except Equation (\ref{meet}) and Equation(\ref{stay}).
\begin{eqnarray} 
 P_{\mathcal{X}'\textrm{joint}}(m\textrm{, }n \textrm{ meet at }x \textrm{ or }y) = 2P^B_{xy}
 \label{meet2}
\end{eqnarray}
\begin{eqnarray} \nonumber
&& P_{\mathcal{X}'\textrm{joint}}(m\textrm{ does not move, } n\textrm{ does not move}) \\
&=& 1 - \sum_{i\in \mathcal{N}_x}P^B_{xi} - \sum_{j\in \mathcal{N}_y}P^B_{yj}.
\label{stay2}
\end{eqnarray}
First, we show that the meeting time of two random walkers following $\mathcal{X}'$ is less than $2
\mathcal{H}_{P^B}{(\mathcal{G})}$. 

For convenience, we adopt the convention of the following notation: if $f(\cdot)$ is a real valued function on the vertex of the graph, then $f(\bar{v})$ is the weighted average of $f(u)$ over all neighbors $u$ of $v$.

Similar as in \cite{Zhu}\cite{Coppersmith}, define a \emph{potential function}
\begin{equation}
\phi(x,y) := \mathcal{H}_{P^B}(x,y) + \mathcal{H}_{P^B}(y, t) - \mathcal{H}_{P^B}(t,y),
\end{equation}
 where $t$ is a hidden vertex on the graph. By Lemma \ref{circle}, $\phi(x,y)$ is symmetric, i.e. $\phi(x,y) = \phi(y,x)$. By the definition of meeting time, $\mathcal{M}$ is also symmetric, i.e. $\mathcal{M}(x,y) = \mathcal{M}(y,x)$. This gives us intuition that we may be able to use $\phi$ to help us bound the meeting time. 
 
 By the definition of hitting time, for $x \neq y$ we have
\begin{eqnarray} \nonumber
&&\mathcal{H}_{P^B}(x, y) \\ \nonumber
&=& 1 + P^B_{xx}\mathcal{H}_{P^B}(x,y) + \sum_{i\in \mathcal{N}_x}P^B_{xi}\mathcal{H}_{P^B}(i,y)\\
&=&1 + w_{xx}\mathcal{H}_{P^B}(x,y) + \sum_{i\in \mathcal{N}_x}w_{xi}\mathcal{H}_{P^B}(i,y), 
\end{eqnarray}
i.e.,
\begin{eqnarray}\nonumber
\mathcal{H}_{P^B}(x, y) &=& \frac{1}{\sum_{i\in \mathcal{N}_x}w_{xi}} + \frac{\sum_{i\in \mathcal{N}_x}w_{xi}\mathcal{H}_{P^B}(i,y)}{\sum_{i\in \mathcal{N}_x}w_{xi}} \\
&=& \frac{1}{\sum_{i\in \mathcal{N}_x}w_{xi}} + \mathcal{H}(\bar{x},y).
\end{eqnarray}
So for $x \neq y$,
\begin{equation} \label{phi}
\phi(x,y) = \frac{1}{\sum_{i\in \mathcal{N}_x}w_{xi}} + \phi(\bar{x},y).
\end{equation}
\begin{eqnarray}\nonumber 
\mathcal{M}_{\mathcal{X}'}(x, y)  &= & 1 + \left( 1 - \sum_{i \in \mathcal{N}_x}P^B_{xi} - \sum_{j \in \mathcal{N}_y}P^B_{yj}\right)\mathcal{M}_{\mathcal{X}'}(x,y) \\ \nonumber 
& + & \sum_{i \in \mathcal{N}_x}P^B_{xi}\mathcal{M}_{\mathcal{X}'}(i,y) \\
& + & \sum_{j \in \mathcal{N}_y}P^B_{yj}\mathcal{M}_{\mathcal{X}'}(x,j).
\label{universal}
\end{eqnarray}
Note that Equation (\ref{universal}) also holds for $x\in \mathcal{N}_y$. We now have
\begin{eqnarray}\nonumber 
&& \left( \sum_{i \in \mathcal{N}_x}P^B_{xi} + \sum_{j \in \mathcal{N}_y}P^B_{yj}\right)\mathcal{M}_{\mathcal{X}'}(x,y)  
\\ & = &  1 +  \sum_{i \in \mathcal{N}_x}P^B_{xi}\mathcal{M}_{\mathcal{X}'}(i,y) 
 +  \sum_{j \in \mathcal{N}_y}P^B_{yj}\mathcal{M}_{\mathcal{X}'}(x,j).
 \label{ineq}
\end{eqnarray}
Equation (\ref{ineq}) shows that at least one of the two inequality below holds:
\begin{equation}
\mathcal{M}_{\mathcal{X}'}(x, y) > \frac{\sum_{i \in \mathcal{N}_x}P^B_{xi}\mathcal{M}_{\mathcal{X}'}(i,y)}{\sum_{i \in \mathcal{N}_x}P^B_{xi}} = \mathcal{M}_{\mathcal{X}'}(\bar{x},y)
\end{equation}
\begin{equation}
\mathcal{M}_{\mathcal{X}'}(x, y) > \frac{\sum_{j \in \mathcal{N}_y}P^B_{yj}\mathcal{M}_{\mathcal{X}'}(x,j)}{\sum_{j \in \mathcal{N}_y}P^B_{yj}} = \mathcal{M}_{\mathcal{X}'}(x,\bar{y})
\label{yineq}
\end{equation}
Without loss of generality, suppose that Equation (\ref{yineq}) holds (otherwise, we can prove the other way round). From Equation (\ref{ineq}), we have
\begin{eqnarray}\nonumber 
&& \sum_{i \in \mathcal{N}_x}P^B_{xi} \mathcal{M}_{\mathcal{X}'}(x,y)  
  =    1 +  \sum_{i \in \mathcal{N}_x}P^B_{xi}\mathcal{M}_{\mathcal{X}'}(i,y) \\
 & + &  \sum_{j \in \mathcal{N}_y}P^B_{yj}\mathcal{M}_{\mathcal{X}'}(x,j) - \sum_{j \in \mathcal{N}_y}P^B_{yj}\mathcal{M}_{\mathcal{X}'}(x,y).
\end{eqnarray}
i.e.,
\begin{eqnarray}\nonumber  
\mathcal{M}_{\mathcal{X}'}(x,y)  
   & = &   \frac{1}{\displaystyle\sum_{i \in \mathcal{N}_x}P^B_{xi}} +  \mathcal{M}_{\mathcal{X}'}(\bar{x},y) \\\nonumber
& + &  \frac{\displaystyle\sum_{j \in \mathcal{N}_y}P^B_{yj}\left(\mathcal{M}_{\mathcal{X}'}(x,\bar{y}) - \mathcal{M}_{\mathcal{X}'}(x,y)\right)}{\displaystyle\sum_{i \in \mathcal{N}_x}P^B_{xi}} \\
& <  & \frac{1}{\displaystyle\sum_{i \in \mathcal{N}_x}w_{xi}} + \mathcal{M}_{\mathcal{X}'}(\bar{x},y).
\label{lastineq}
\end{eqnarray}
Now we claim that $\mathcal{M}_{\mathcal{X}'}(x,y) \le \phi(x,y)$. Suppose it is not the case. Let $\beta = \max_{x,y}\{\mathcal{M}_{\mathcal{X}'}(x,y) - \phi(x,y)\}$. Among all the pairs $x, y$ realizing $\beta$, choose any pair.
It is clear that $x \neq y$, since $\mathcal{M}_{\mathcal{X}'}(x,x) = 0 \le \phi(x,x)$. By Equation (\ref{phi}) and Equation (\ref{lastineq}), 
\begin{eqnarray}\nonumber
\mathcal{M}_{\mathcal{X}'}(x,y) &=& \phi(x,y) + \beta \\ \nonumber
&=& \frac{1}{\sum_{i\in \mathcal{N}_x}w_{xi}} + \phi(\bar{x}, y) + \beta\\  \nonumber
&\ge& \frac{1}{\sum_{i\in \mathcal{N}_x}w_{xi}} + \mathcal{M}_{\mathcal{X}'}(\bar{x},y) \\
& > &\mathcal{M}_{\mathcal{X}'}(x,y).
\end{eqnarray}
Contradiction. Thus $\mathcal{M}_{\mathcal{X}'}(\mathcal{G}) <  \phi(x,y) < 2\mathcal{H}_{P^B}{(\mathcal{G})}$.

Now we are ready to proof theorem \ref{meeting}. Compare the joint distribution of $\mathcal{X}$ and $\mathcal{X'}$, we notice that the two two-dimensional Markov chains are coupled until two random walkers meet, because half of the time when the two random walkers in $\mathcal{X}'$ meet, the random walkers in  $\mathcal{X}$ do not, but stay in the same position. We claim that $\mathcal{M}_{\mathcal{X}}(\mathcal{G}) \leq 2\mathcal{M}_{\mathcal{X}'}(\mathcal{G})$. 

In the random process $\mathcal{X}'$, when two random walkers $m$, $n$ meet, instead of finishing the process, we let them cross and keep doing random walk according to $P_{\mathcal{X}'\textrm{joint}}$. The expected length of each cross is less than or equal to $\mathcal{M}_{\mathcal{X}'(\mathcal{G})}$. At each cross, the random process $\mathcal{X}$ finishes with a probability of 1/2, independently. Thus for any $x, y \in \mathcal{V}$ we have 
\begin{equation}
\mathcal{M}_{\mathcal{X}}(x,y) \leq \sum_{i = 1}^{\infty}\left(\frac{1}{2}\right)^ii \mathcal{M}_{\mathcal{X}'}(\mathcal{G}) = 2\mathcal{M}_{\mathcal{X}'}(\mathcal{G}).
\end{equation}

This completes the proof.
\end{proof}

\subsection{Convergence Speed Analysis}
Without loss of generality, let us suppose that in the initial setting, more nodes hold \emph{strong positive} opinions ($S^+$). As briefly analyzed in Section \ref{def}, the process undergoes two stages: the depletion of $S^-$ and the depletion of $W^-$. By our assumption, 
\begin{equation}
\label{ }
|S^+(0)| > |S^-(0)|
\end{equation}
and 
\begin{equation}
\label{ }
|S^+(0)| + |S^-(0)| = N,
\end{equation} 
where $N$ is the number of nodes on the graph.  

According to the update rules in Section \ref{sec:problem}, we have
\begin{equation}
\label{ }
|S^+(t)| - |S^-(t)| = |S^+(0)|-|S^-(0)|.
\end{equation}

Let $T_1$ and $T_2$ denote the maximum expected time it takes for Stage 1 and Stage 2 to finish. In the first stage, two opposite strong opinions annihilate when an edge between them is activated. Otherwise they take biased random walk on the graph $\mathcal{G}$.  
In the second stage, the remaining $|S^+(0)|-|S^-(0)|$ \emph{Strong Positive}s take random walks over graph $\mathcal{G}$, transforming \emph{Weak Negative} into \emph{Weak Positive}. 
Let $CT_{\mathcal{G}}(v)$ denote the expected time for a random walker starting from node $v$ to meet all other random walkers who are also taking random walks on the same graph but starting from different nodes. Define 
$$CT(\mathcal{G}) = \max_{v \in \mathcal{V}} CT_{\mathcal{G}}(v).$$

\begin{cor}\label{max}
\begin{equation}
\label{ }
CT(\mathcal{G}) = \mathcal{O}(N^4\log N).
\end{equation}\end{cor}
\begin{proof}
In terms of the global clock ticks $\{Z_k\}_{k\ge0}$ in our setting, by Theorem \ref{hitting} and Theorem \ref{meeting}, we have  
\begin{equation}
\label{ }
\mathcal{M}_{(\mathcal{G})} < 2\mathcal{H}_{(\mathcal{G},P^B)} < N^4. 
\end{equation}
Thus a union bound for $CT(\mathcal{G})$ is $\sim \frac{1}{2}N^5$ is available since there are no more than $\frac{N} {2}$ consecutive meetings.

In order to obtain a tighter bound for $CT(\mathcal{G})$, we divide the random walks into $\ln N$ periods of length $k\mathcal{M}(\mathcal{G})$ each, where $k$ is a constant. Let $a$ be the ``special" random walker trying to meet all other random walkers. For any period $i$ and any other random walker $v$, by the Markov inequality, we have
\begin{eqnarray}\nonumber
&&\Pr(\textrm{$a$ does not meet $v$ during period $i$})  \\ \nonumber
&\le& \frac{\mathcal{M}_{\mathcal{X}'}(\mathcal{G})}{k\mathcal{M}_{\mathcal{X}'}(\mathcal{G})} \\
&=& \frac{1}{k}
\end{eqnarray}
so
\begin{eqnarray}\nonumber
&&\Pr(\textrm{$a$ does not meet $v$ during any period})  \\
&\le& \left( \frac{1}{k}\right)^{\ln N} = N^{-\ln k}
\end{eqnarray}
If we take the union bound, 
\begin{eqnarray}\nonumber
&&\Pr(\textrm{$a$ doesn't meet some walker during any period}) \\
 &\le&  N\cdot N^{-\ln k}.
\end{eqnarray}

Conditioning on whether or not the walker $a$ has met all other walkers after all $k\mathcal{M}(g)\ln N$ steps, and using the previous $\mathcal{O}(N^5)$ upper bound, we have

\begin{eqnarray}\nonumber
CT(\mathcal{G}) &\le& k\mathcal{M}(g)\ln N + N\cdot N^{-\ln k}\frac{1}{2}N^5\\ 
&=& k\mathcal{M}(g)\ln N +  \frac{1}{2}N^{6-\ln k}
\end{eqnarray}

When $k$ is sufficiently large, say $k \ge e^6$, the second term is small. So
\begin{equation}
\label{ }
CT(\mathcal{G}) \le cN^4\ln N.
\end{equation}
This completes the proof. 
\end{proof}

\begin{cor}\label{c2}
\begin{equation}
\label{t1}
\textrm{Stage 1:	}T_1  \le 2CT(\mathcal{G})
\end{equation}
\begin{equation}
\label{t2}
\textrm{Stage 2:	}T_2  \le CT(\mathcal{G})
\end{equation}
\end{cor}
\begin{proof}
In order to prove (\ref{t1}), we can construct a coupling process of stage 1. The coupling process is when two different strong opinions meet, instead of following the rules to change into weak opinions, they just keep their states and keep moving along the same path they would have as weak opinions. This process is over when every strong opinion has met all other opposite strong opinions, by when Stage 1 must have finished, i.e. before at most $N^2/4$ such meetings. The rest of the proof just follows from Corollary \ref{max}, except we divide the random walks into $\ln(N^2)$ periods of length $k\mathcal{M}(\mathcal{G})$ instead of $\ln N$.
Equation (\ref{t2}) follows from the fact that there are at most $N-1$ meetings which a single strong opinion meets all the weak opinions to make sure convergence. 
\end{proof}

Theorem \ref{mainth} is then a direct result from Corollary \ref{max} and Corollary \ref{c2}. 

\section{Simulation Results}
In this section, we give a simple example of star networks in order to show how to use the analysis in Section \ref{main} for the particular graph with known topology. Simulation results are provided to validate the analysis. 
We also simulate the distributed process on Erd\"os-R\'enyi random graph in order to get an insight on how the algorithm performs on a random graph.


\subsection{Star Networks}
\begin{figure}[!t]
\centering
\centerline{\includegraphics[width=3.5cm]{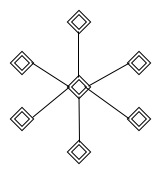}}
\caption{{\em A star network.}}
\label{star}
\end{figure}

Star networks are one of the most common network topologies. A star network $\mathcal{S}$ of $N$ nodes has one central hub and $N-1$ leaf nodes, as shown in Fig. \ref{star}. Now let us derive an upper bound following the similar analysis in Section \ref{main}.

\subsubsection{Analysis}\label{starbound}
By Equation (\ref{weight}) in Section \ref{mt}, suppose that there is a star network of $N$ nodes, with the central hub denoted as $c$. For $\forall i,j \ne c$, we have
\begin{equation}
w_{ic} = w_{jc} = \frac{1}{N}\left(1 + \frac{1}{N-1}\right) = \frac{1}{N-1}.
\end{equation}
The equivalent resistance between any two leaf nodes $i$ and $j$ is
\begin{equation}
r'_{ij} = \frac{1}{w_{ic}} + \frac{1}{w_{jc}} = 2N - 2.
\end{equation}   
By the symmetry of the star network, it is easy to see that
\begin{equation}
\mathcal{H}_{P^B}(i, j) = \mathcal{H}_{P^B}(j, i).
\end{equation}
By Equation (\ref{commute}),
\begin{equation}
\mathcal{H}_{P^B}(\mathcal{S}) = \mathcal{H}_{P^B}(i,j) = N(N-1).
\end{equation}
Then following Theorem 2 and similar analysis of Corollary 1, 2 and 3, we can bound the convergence time of a star network by $O(N^2\log N)$. 
\subsubsection{Simulations} \label{sim1}
In order to  justify the bound we derived in Section \ref{starbound}, we did simulations on star networks with nodes $N$ ranging from 21 to 481, with an interval of 20. Initially, there are $ceil(N)$ \emph{strong positive} and $floor(N)$ \emph{strong negative} nodes, i.e., $|S^+| - |S^-| = 1$. Those nodes communicate with each other following the protocol in Section \ref{sec:problem}. The process finishes when consensus is reached. Convergence time is the average of 20 rounds of simulations. 

\begin{figure}[t]
\centering
\centerline{\includegraphics[width=.45\textwidth]{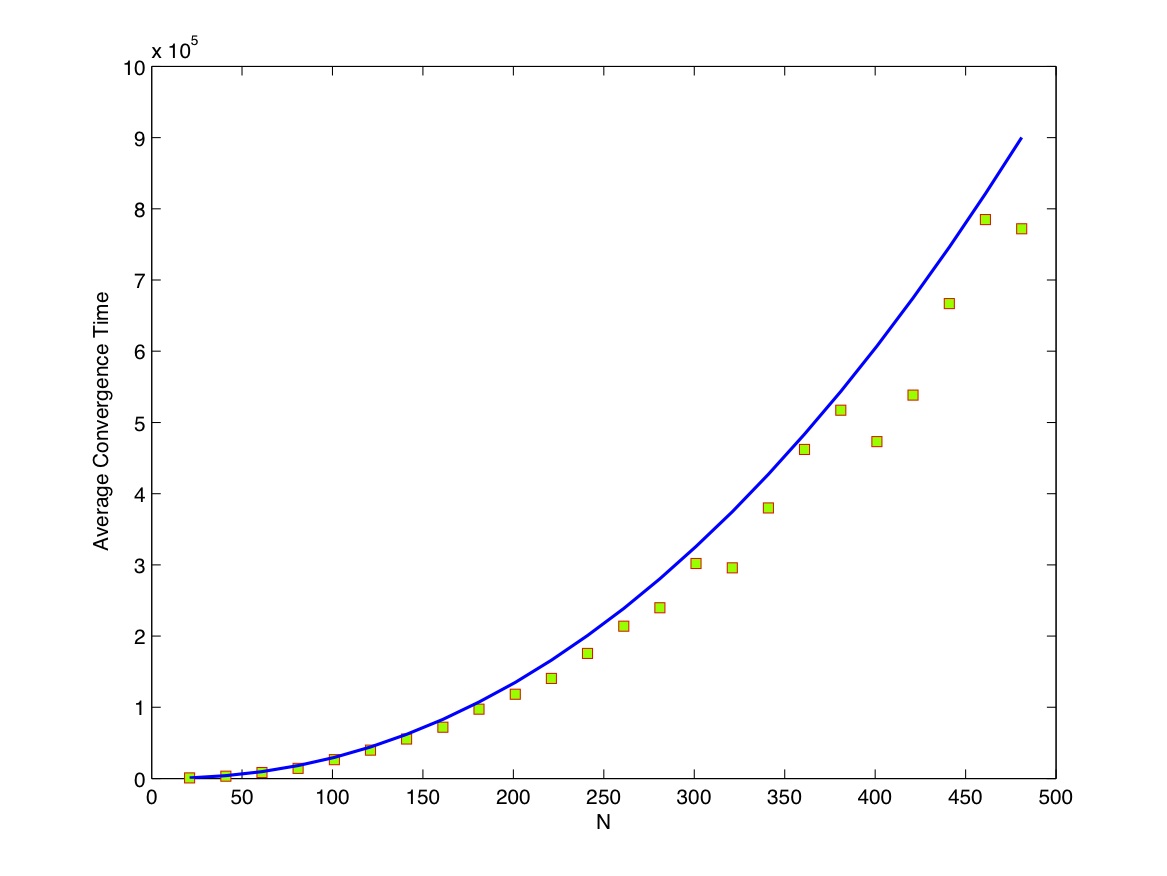}}
\caption{{\em Average convergence time (green squares) versus the size of the star network. The blue solid line indicate $0.63N^2\log N$. }}
\label{simulation}
\end{figure}

We plot the average convergence time (green squares) versus the size of star networks in Fig. \ref{simulation}. As in the figure indicated by the solid blue line, $O(N^2\log N)$ is indeed an upper bound. Fig. \ref{simulation} also indicates that this bound  is tight for a star network.


%


%

\subsection{Erd\"os-R\'enyi random graph}
In a Erd\"os-R\'enyi random graph $\mathcal{R}$, an edge is set between each pair of nodes independently with equal probability $p$. As one of the properties of Erd\"os-R\'enyi random graphs, when $p > \frac{(1+\epsilon)\log N}{N}$, the graph $\mathcal{R}$ will almost surely be connected \cite{Durrett}. $$E(\textrm{number of edges}) = 0.5N(N-1)p.$$ The diameter of Erd\"os-R\'enyi random graphs is rather sensitive to small changes in the graph, but the typical distance between two random nodes on the graph is
$d = \frac{\log N}{\log (pN)}$\cite{Durrett}.

\begin{figure}[!t]
\centering
\centerline{\includegraphics[width=.45\textwidth]{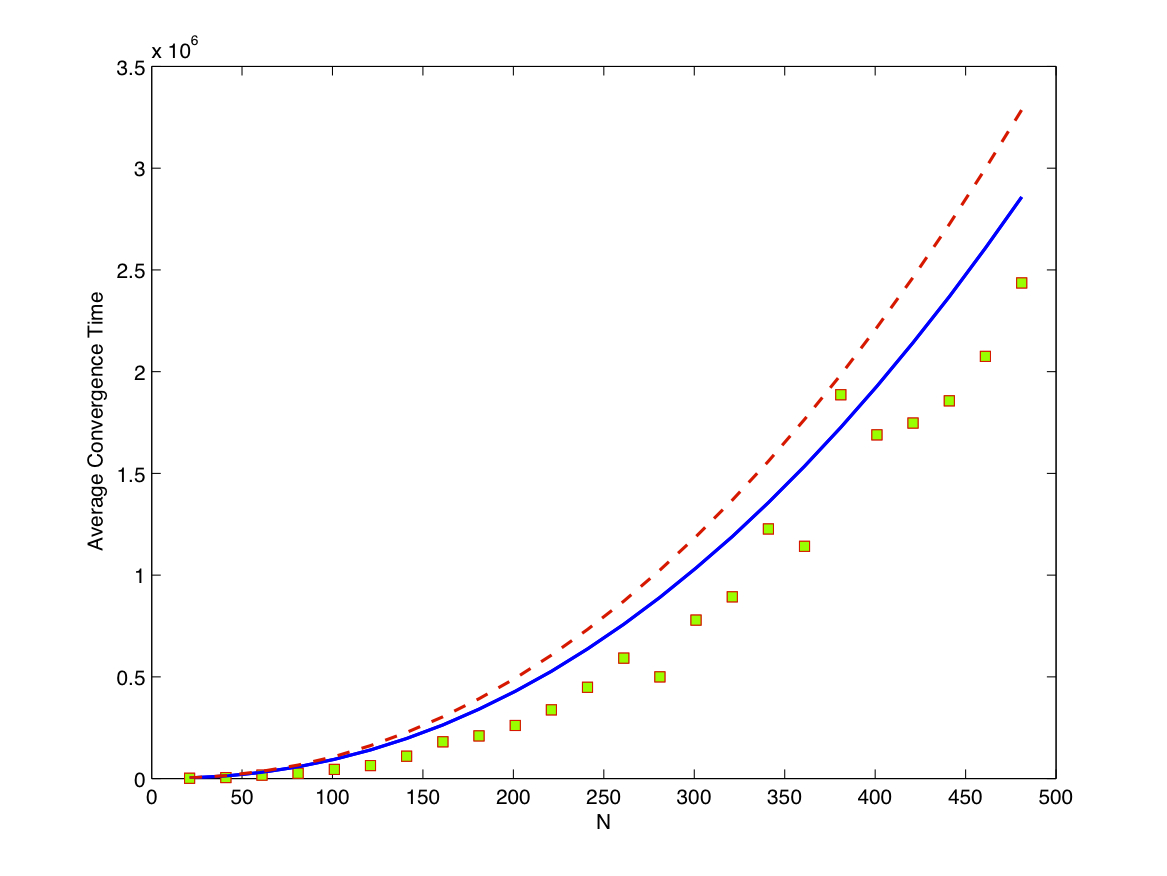}}
\caption{{\em Average convergence time versus the size of the Erd\"os-R\'enyi random graph. The blue solid line indicates $2N^2\log N$, and the red dash line indicates $2.3N^2\log N$.}}
\label{simulation2}
\end{figure}

We created Erd\"os-R\'enyi random graph by setting $p = 5\log N /N$, where $N$ ranged from $21$ to $481$, with an interval of $20$. Other settings are the same as in Section \ref{sim1}. We plot  the average convergence time (green squares) versus the size of Erd\"os-R\'enyi random graph in Fig. \ref{simulation2}.  As indicated in the figure, the expected convergence time is in the order of $N^2\log N$.
%

%

%

\section{conclusions}
\label{sec: discussion}
In this paper, we use the theory of electric network, random walks, and couplings of Markov chains to derive a polynomial bound on convergence time with respect to the size of the network, for the distributed binary consensus problem addressed in \cite{Benezit}. It can be extended to a tighter bound for any given topology of network using the effective resistance analogy.  Our result provides insights of the performance of the binary consensus algorithm, and with applications in sensor networks, distributed computing, peer-to-peer systems, etc..

In the analysis in Section \ref{main}, we notice that if some sensors do not have an observation, we can randomly assign weak opinions to them. Doing this does not change the consensus result if there is a majority opinion at the beginning. 

\section{acknowledgement}
This research was supported in part by the Center for Science of
Information (CSoI), an National Science Foundation (NSF) Science and Technology Center, under grant
agreement CCF-0939370, by NSF under the grant CCF-1116013,
by the U.S. Army Research Office under grant
number W911NF-07-1-0185, and by a research grant from Deutsche Telekom
AG.


\bibliographystyle{IEEEbib}
\bibliography{myrefs}

\end{document}